\documentclass[aps,twocolumn,longbibliography]{revtex4-1}

\usepackage{amsmath}
\usepackage{amsfonts}
\usepackage{amssymb}
\usepackage{braket}
\usepackage[version=3]{mhchem}
\usepackage{graphicx}
\usepackage{dcolumn}
\usepackage{bm}

\usepackage{txfonts}
\usepackage[usenames,dvipsnames]{xcolor}
\usepackage{array}

\begin{document}


\title{Kitaev-Heisenberg Hamiltonian for high-spin $d^7$ Mott insulators}
 

\author{Ryoya Sano, Yasuyuki Kato, and Yukitoshi Motome}
\affiliation{Department of Applied Physics, The University of Tokyo, Tokyo 113-8656, Japan}


\date{\today}

\begin{abstract}
In the search for quantum spin liquids, candidate materials for the Kitaev model and its extensions have been intensively explored during the past decade, as the models realize the exact quantum spin liquids in the ground state. 
Thus far, insulating magnets in the low-spin $d^5$ electron configuration under the strong spin-orbit coupling have been studied for realizing the Kitaev-type bond-dependent anisotropic interactions between the spin-orbital entangled Kramers doublets. 
To extend the candidates, here we investigate the systems in a high-spin $d^7$ electron configuration, whose ground state is described by the spin-orbital entangled Kramers doublet. 
By the second-order perturbation in terms of the $t_{2g}$-$t_{2g}$ and $t_{2g}$-$e_g$ hoppings, we show that the effective spin model possesses the anisotropic Kitaev interactions as well as the isotropic Heisenberg ones. 
While the Kitaev interaction is always ferromagnetic, the Heisenberg interaction can become either ferromagnetic or antiferromagnetic depending on the Coulomb interactions and the crystalline electric fields. 
We also derive the effective model for the low-spin $d^5$ electron configuration within the same perturbation scheme, in which the Kitaev interaction becomes both ferromagnetic and antiferromagnetic, while the Heisenberg one always ferromagnetic. 
Referring to the previous study for the Kitaev-Heisenberg model, we find that the quantum spin liquid phase exists in the reasonable parameter region in both $d^7$ and $d^5$ cases, while the former has a richer structure of the phase diagram. 
We discuss the advantages of the $d^7$ case in comparison with the $d^5$ case. 
Our results indicate that the high-spin $d^7$ state provides another platform for the Kitaev-type quantum spin liquid.
\end{abstract}


\maketitle


\section{Introduction}
\label{sec:int}

The quantum spin liquid (QSL) is an exotic state of matter in insulating magnets~\cite{ANDERSON1973153,balents2010spin,0034-4885-80-1-016502, RevModPhys.89.025003}. 
They do not  show any long-range order in the sense of the Landau theory down to zero temperature, whereas they may host a long-range quantum entanglement. 
The exotic states have attracted much attention because of the unusual properties, such as the fractionalization of spin degrees of freedom~\cite{sachdev1992kagome} and emergent anyonic quasiparticles in two-dimensional systems~\cite{PhysRevB.43.10289, Kitaev20062}. 
 
The Kitaev model is a canonical model providing such QSLs in the exact ground states~\cite{Kitaev20062}.
It is a quantum spin $S=1/2$ model defined on a honeycomb lattice, whose Hamiltonian is given by 
\begin{equation}
\mathcal{H}_{\mathrm {Kitaev}}=
-J_x\sum_{\langle i,j \rangle_x}S^x_iS^x_j
-J_y\sum_{\langle i,j \rangle_y}S^y_iS^y_j
-J_z\sum_{\langle i,j \rangle_z}S^z_iS^z_j.
\label{eq:Kitaev_Hamiltonian}
\end{equation}
Here, each sum is taken for the nearest-neighbor $\mu$ bonds on the honeycomb lattice, where $\mu=x,y,z$ represents one of three different bond directions on the tri-coordinate lattice; $S^{\mu}_i$ denotes the $\mu$ component of $S=1/2$ spin operator at site $i$, and $J_\mu$ is the coupling constant. 
The Hamiltonian consists of the Ising-type anisotropic interactions whose spin components depend on the bond directions.  
Remarkably, the ground state of this quantum spin model is exactly obtained because the model has macroscopic number of the constants of motion and can be mapped to a free Majorana fermion problem~\cite{Kitaev20062}. 
The exact ground state is shown to be a QSL with extremely short-range spin correlations (nonzero only for the nearest neighbors as well as the same sites)~\cite{PhysRevLett.98.247201}. 
In the QSL, a quantum spin $S=1/2$ is fractionalized into emergent quasiparticles: itinerant Majorana fermions and static $Z_2$ fluxes~\cite{Kitaev20062}.

The Kitaev-type bond-dependent anisotropic interactions in Eq.~(\ref{eq:Kitaev_Hamiltonian}) are argued to potentially realize in materials with strong entanglement between the spin and orbital degrees of freedom~\cite{PhysRevLett.102.017205}. 
Such interesting possibility was first suggested in Ref.~\cite{PTPS.160.155}. 
Suppose the atomic state is in the low-spin $d^5$ electron configuration in the $t_{2g}$ manifold under the octahedral crystalline electric field (CEF) with the strong spin-orbit coupling (SOC), the lowest-energy multiplet is given by the spin-orbital entangled Kramers doublet, denoted by the pseudospin $J_{\mathrm{eff}}=1/2$. 
When such octahedra share their edges, the second-order perturbation in terms of hopping processes via the neighboring ligand ions leads to the Kitaev-type interaction between the pseudospin $J_{\mathrm{eff}}=1/2$ moments, because of quantum interference between different perturbation processes~\cite{PhysRevLett.102.017205}.   
In reality, other interactions, such as the Heisenberg exchange interaction and further-neighbor interactions, coexist with the Kitaev interactions. 
Such a situation is described by extended Kitaev models~\cite{PhysRevLett.105.027204, PhysRevLett.110.097204,PhysRevLett.112.077204,PhysRevB.90.155126,PhysRevB.91.241110,PhysRevB.94.064435}, which have been intensively studied for understanding of the candidate materials, such as Ir and Ru compounds~\cite{PhysRevB.82.064412,PhysRevLett.108.127203,PhysRevLett.109.266406,modic2014realization,PhysRevB.90.041112,PhysRevB.91.094422,chun2015direct}.

Thus, the necessary conditions for the Kitaev interactions are 
(i) the effective pseudospin degree of freedom arising from the spin-orbital entanglement by the strong SOC and 
(ii) the orbital-dependent hopping processes suffering from quantum interference. 
These ingredients are not necessarily limited to the low-spin $d^5$ electron configuration with edge-sharing octahedra.  
Nevertheless, most of the previous studies were restricted to the $d^5$ compounds, 
except for a recent attempt for rare-earth materials~\cite{PhysRevB.95.085132}.
For extending the material quest for the Kitaev-type QSL, it is intriguing to pursuit the possibility of the Kitaev interactions in other electron configurations.
  
In this paper, we theoretically propose another situation potentially relevant to a realization of the Kitaev-type interaction. 
We consider the high-spin $d^7$ electron configuration under the octahedral CEF and the strong SOC, which results in the spin-orbital entangled Kramers doublet~\cite{griffith1961theory}. 
Regarding this doublet as the pseudospin degree of freedom, we derive an effective Hamiltonian in the strong coupling limit by the second-order perturbation theory in terms of the $d$-$p$-$d$ hoppings for the edge-sharing configuration similar to that considered for the low-spin $d^5$ case.
We find that the effective Hamiltonian includes both the Kitaev-type bond-dependent anisotropic interaction and the isotropic Heisenberg interaction, called the Kitaev-Heisenberg model~\cite{PhysRevLett.105.027204}. 
We show that the coupling constants depend on the Coulomb interactions and the crystalline electric fields from ligand ions, and interestingly, the Heisenberg interaction can become both ferromagnetic and antiferromagnetic, while the Kitaev one is always ferromagnetic. 
We also derive an effective model for the low-spin $d^5$ case within the similar perturbation scheme; in this case, the Kitaev interaction can be both ferromagnetic and antiferromagnetic, while the Heisenberg one is always ferromagnetic. 
By mapping the results onto those in the literature~\cite{PhysRevLett.105.027204,PhysRevLett.110.097204}, we show the ground-state phase diagrams for both the $d^7$ and $d^5$ models. 
In both cases, we find the QSL phase in a reasonable parameter region, while the $d^7$ case has a richer structure because of the Heisenberg interaction changing the sign.
We discuss the advantages of the high-spin $d^7$ case for the realization of Kitaev QSL in comparison with the low-spin $d^5$ case. 

The organization of the rest of this paper is as follows. 
In Sec.~\ref{sec:model}, we describe the derivation of the low-energy effective Hamiltonian for the $d^7$ situation, starting from the multi-orbital Hubbard Hamiltonian. 
In Sec.~\ref{sec:result}, we show the explicit form of the effective Hamiltonian and elucidate the ground-state phase diagram. 
We also discuss the $d^5$ case for comparison. 
In Sec.~\ref{dis}, we compare the results for the $d^7$ and $d^5$ cases and discuss the advantages of the former. 
Section~\ref{sum} is devoted to the summary.

\section{Model and method}
\label{sec:model}

\subsection{Multi-orbital Hubbard model}
\label{subsec:multi}

Following the previous study for the low-spin $d^5$ electron configuration in Ref.~\cite{PhysRevLett.102.017205}, we consider the edge-sharing network of $MX_6$ octahedra ($M$ and $X$ are a transition metal and a ligand, respectively) comprising a honeycomb network of $M$ cations. 
To derive a low-energy effective Hamiltonian for $d$ electrons of $M$ cations on the honeycomb network, we begin with a multi-orbital Hubbard model for the $d$ electron manifold. 
The Hamiltonian is composed of four terms as
\begin{equation}
\mathcal{H}=
\mathcal{H}_{{\mathrm{CEF}}}+
\mathcal{H}_{\mathrm{int}}+
\mathcal{H}_{\mathrm{SOC}}
+\mathcal{H}_{\mathrm{hop}},
\label{eq:H_multi}
\end{equation} 
where each term describes the CEF, Coulomb interactions between $d$ electrons, SOC, and electron hopping.

\begin{figure}[htp]
\centering
\includegraphics[width=0.85\columnwidth,clip]{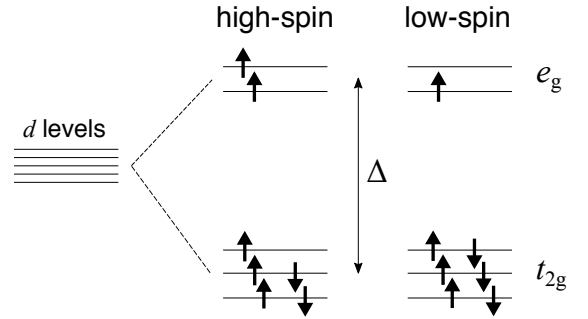}
\caption{\label{fig:level}
Atomic $d$ levels splitting into two groups under the octahedral CEF $\Delta$:  $e_g$ levels at a higher energy and $t_{2g}$ levels at a lower energy. 
The $d^7$ electron configuration can take either high-spin (middle) or low-spin state (right), depending on the strength of Coulomb interactions and $\Delta$. 
}
\end{figure}

For the first term $\mathcal{H}_{\mathrm{CEF}}$ in Eq.~(\ref{eq:H_multi}), we take into account the octahedral CEF, which splits $d$ levels into the $e_g$ and $t_{2g}$ manifolds (see Fig.~\ref{fig:level}). 
The higher-energy $e_g$ manifold is composed of $d_{3z^2-r^2}$ and $d_{x^2-y^2}$ orbitals, while the lower-energy $t_{2g}$ manifold is composed of $d_{yz}$, $d_{zx}$, and $d_{xy}$. 
Then, the CEF term is given in the form
\begin{eqnarray}
\mathcal{H}_{\mathrm{CEF}}=\Delta \sum_{i \sigma}(n_{iu \sigma}+n_{iv \sigma}),
\label{eq:H_CEF}
\end{eqnarray}
where $\Delta >0$ denotes the energy difference between the $e_g$ and $t_{2g}$ manifolds, and we set the energy for the $t_{2g}$ manifold at zero; 
$u$ and $v$ denote the $d_{3z^2-r^2}$ and $d_{x^2-y^2}$ orbitals, respectively (we denote $d_{yz}$, $d_{zx}$, and $d_{xy}$ by $\xi$, $\eta$, and $\zeta$, respectively, in the following). 
$n_{i\alpha \sigma}$ denotes the number operator for electrons in orbital $\alpha$ with spin $\sigma$ at site $i$.

For the interaction term $\mathcal{H}_{\mathrm{int}}$ in Eq.~(\ref{eq:H_multi}), we take into account the on-site Coulomb interaction, which is given in the form
\begin{align}
\mathcal{H}_{\mathrm{int}}&=U\sum_{i\alpha}n_{i\alpha \uparrow}n_{i\alpha \downarrow}
+\frac{U'}{2}\sum_{i,\alpha,\beta(\alpha  \neq  \beta)}n_{i\alpha}n_{i\beta}\nonumber\\
&-\frac{J}{2}
\sum_{i,\sigma,\sigma',\alpha,\beta(\alpha \neq \beta)}c^{\dagger}_{i\alpha \sigma}c^{\;}_{i\alpha \sigma'}c^{\dagger}_{i\beta \sigma'}c^{\;}_{i\beta \sigma}\nonumber\\
&-\frac{J'}{2}
\sum_{i,\sigma,\sigma',\alpha,\beta(\sigma \neq \sigma',\alpha \neq \beta)}c^{\dagger}_{i\alpha \sigma}c^{\;}_{i\beta \sigma'}c^{\dagger}_{i\alpha \sigma'}c^{\;}_{i\beta \sigma},
\label{eq:H_int}
\end{align}
where $n_{i \alpha } = \sum_{\sigma} n_{i \alpha \sigma}$, and $c_{i \alpha \sigma}$ ($c^\dagger_{i \alpha \sigma}$) is the annihilation (creation) operator of an electron in orbital $\alpha$ with spin $\sigma$ at site $i$. 
Here, $U$, $U'$, $J$, and $J'$ denote the intra-orbital Coulomb repulsion, the inter-orbital Coulomb repulsion, the Hund's-rule coupling, and the pair-hopping interaction, respectively ($U$, $U'$, $J$, and $J'$ are all taken to be positive). 
As all the bases $\xi$, $\eta$, $\zeta$, $u$, and $v$ are described by the real wave functions, the relation $J=J'$ holds. 
For simplicity, we assume the spherical rotational symmetry for the $d$ orbitals, which assures the relation $U'=U-2J$. 

In the following, we focus on the case in which each transition metal cation has seven $d$ electrons on average ($d^7$ state), while in Sec.~\ref{subsec:d5} we also revisit the case with five $d$ electrons studied previously~\cite{PhysRevLett.102.017205}. 
In the atomic limit, i.e., $\mathcal{H}_{\mathrm{hop}} = 0$ in Eq.~(\ref{eq:H_multi}), the $d^7$ state can take either high or low-spin configuration, depending on the parameters $U$, $J$, and $\Delta$.
The high-spin state (12-fold degenerate) is represented as $t_{2g}^5e_{g}^{2}({}^3A_{2g}){}^{4}T_{1g}$ (see the middle panel of Fig.~\ref{fig:level}). 
Meanwhile, the low-spin state (fourfold degenerate) is represented as $t_{2g}^6e_{g}^{1}\, {}^{2}E_{g}$ (see the right panel of Fig.~\ref{fig:level}).  
In this study, we consider the situation that the high-spin state has a lower energy than the low-spin state, as the SOC is ineffective for the latter. 
Such a situation is realized under the conditions 
\begin{eqnarray}
	U > 3J > \frac{3}{\sqrt{22}-1}\Delta.
	\label{eq:high-spin conditions}
\end{eqnarray}

When we restrict ourselves to the high-spin 12-fold manifold, $\mathcal{H}_{\mathrm{SOC}}$ in Eq.~(\ref{eq:H_multi}) is given in the form
\begin{eqnarray}
\mathcal{H}_{\mathrm{SOC}}= \lambda \sum_i  \tilde{\bf L}_{i} \cdot \tilde{\bf S}_{i}, 
\label{eq:H_SOC}
\end{eqnarray} 
where $\lambda>0$ denotes the SOC constant,
$\tilde{\bf L}_{i}$ denotes the fictitious orbital angular momentum operator for the basis of ${}^{4}T_{1g}$ representation at site $i$~\cite{sugano2012multiplets}, and $\tilde{\bf S}_{i}=\mathcal{P}\sum_{\alpha}{\bf s}_{i\alpha}\mathcal{P}^\dagger$ denotes the spin angular momentum operator of $S=3/2$ at site $i$ (${\bf s}_{i\alpha}$ denotes the spin operator for an electron in the $\alpha$ orbital at site $i$ and $\mathcal{P}$ is a projection operator onto the Hilbert space of $S=3/2$). 
The matrix form of $\tilde{L}$ for the basis for $T_{1g}$ irreducible representation is given by
\begin{align}
	\tilde{L}^x =
  \left(
  \ket{x}\ket{y}\ket{z}
  \right)
   \left(
    \begin{array}{ccc}
      0 & 0 & 0 \\
      0 & 0 & -i \\
      0 & i & 0
    \end{array}
  \right)
  \left(
        \begin{array}{c}
        \bra{x}\\
        \bra{y}\\
        \bra{z}
    	\end{array}
    	  \right),\\
  \tilde{L}^y =
    \left(
    \ket{x}\ket{y}\ket{z}
    \right)
   \left(
    \begin{array}{ccc}
      0 & 0 & i \\
      0 & 0 & 0 \\
      -i & 0 & 0
    \end{array}
  \right)
  \left(
          \begin{array}{c}
          \bra{x}\\
          \bra{y}\\
          \bra{z}
      	\end{array}
      	  \right),\\
  \tilde{L}^z = 
    \left(
    \ket{x}\ket{y}\ket{z}
    \right)
  \left(
    \begin{array}{ccc}
      0 & -i & 0 \\
      i & 0 & 0 \\
      0 & 0 & 0
    \end{array}
  \right)
  \left(
          \begin{array}{c}
          \bra{x}\\
          \bra{y}\\
          \bra{z}
      	\end{array}
      	  \right).
\end{align}
The SOC acts on the Hilbert space spanned by \{$\ket{\tilde{S}^z,\mu}=\ket{\tilde{S}^z}\ket{\mu}$\}. 
$\ket{\tilde{S}^z,\mu}$ represents a basis for the $t_{2g}^5e_{g}^{2}({}^3A_{2g}) {}^{4}T_{1g}$ irreducible representation with the $z$ component of $\tilde{{\bf S}}$, $\tilde{S}^z$, and $\mu=x,y,z$: each basis  is explicitly given as 
\begin{align}
\Ket{\frac{3}{2},x}
&=c^{\dagger}_{\xi \uparrow}c^{\dagger}_{\eta \uparrow}c^{\dagger}_{\eta \downarrow}c^{\dagger}_{\zeta \uparrow}c^{\dagger}_{\zeta \downarrow}c^{\dagger}_{u \uparrow}c^{\dagger}_{v \uparrow}\ket{0},
\label{eq:Sz_3x/2}
\\
\Ket{\frac{1}{2},x}
&=\frac{1}{\sqrt{3}}c^{\dagger}_{\xi \uparrow}c^{\dagger}_{\eta \uparrow}c^{\dagger}_{\eta \downarrow}c^{\dagger}_{\zeta \uparrow}c^{\dagger}_{\zeta \downarrow}(c^{\dagger}_{u \uparrow}c^{\dagger}_{v \downarrow}+c^{\dagger}_{u \downarrow}c^{\dagger}_{v \uparrow})\ket{0}\nonumber \\
&+\frac{1}{\sqrt{3}}c^{\dagger}_{\xi \downarrow}c^{\dagger}_{\eta \uparrow}c^{\dagger}_{\eta \downarrow}c^{\dagger}_{\zeta \uparrow}c^{\dagger}_{\zeta \downarrow}c^{\dagger}_{u \uparrow}c^{\dagger}_{v \uparrow}\ket{0},
\label{eq:Sz_x/2}
\\
\Ket{-\frac{1}{2},x}
&=\frac{1}{\sqrt{3}}c^{\dagger}_{\xi \downarrow}c^{\dagger}_{\eta \uparrow}c^{\dagger}_{\eta \downarrow}c^{\dagger}_{\zeta \uparrow}c^{\dagger}_{\zeta \downarrow}(c^{\dagger}_{u \uparrow}c^{\dagger}_{v \downarrow}+c^{\dagger}_{u \downarrow}c^{\dagger}_{v \uparrow})\ket{0}\nonumber \\
&+\frac{1}{\sqrt{3}}c^{\dagger}_{\xi \uparrow}c^{\dagger}_{\eta \uparrow}c^{\dagger}_{\eta \downarrow}c^{\dagger}_{\zeta \uparrow}c^{\dagger}_{\zeta \downarrow}c^{\dagger}_{u \downarrow}c^{\dagger}_{v \downarrow}\ket{0},
\label{eq:Sz_-x/2}
\\
\Ket{-\frac{3}{2},x}
&=c^{\dagger}_{\xi \downarrow}c^{\dagger}_{\eta \uparrow}c^{\dagger}_{\eta \downarrow}c^{\dagger}_{\zeta \uparrow}c^{\dagger}_{\zeta \downarrow}c^{\dagger}_{u \downarrow}c^{\dagger}_{v \downarrow}\ket{0},
\label{eq:Sz_-3x/2}
\end{align}
where $\ket{0}$ means the vacuum of $d$ electrons, and the states such as $\ket{\tilde{S}^z,y}$ and $\ket{\tilde{S}^z,z}$ can be obtained by cyclic permutations of $\xi$, $\eta$, and $\zeta$ in Eqs.~(\ref{eq:Sz_3x/2})-(\ref{eq:Sz_-3x/2}), i.e., \{$\xi \eta \zeta$\} $\rightarrow$ \{$\eta \zeta \xi$\} and \{$\zeta \xi \eta$\} for $\ket{\tilde{S}^z, y}$ and $\ket{\tilde{S}^z, z}$, respectively.

\subsection{Kramers doublet}
\label{subsec:doublet}

\begin{figure}[htp]
\centering
\includegraphics[width=0.9\columnwidth,clip]{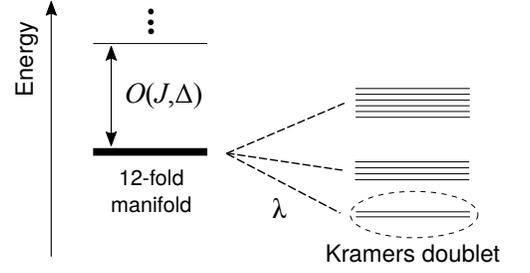}
\caption{\label{fig:split} 
Schematic figure for the formation of Kramers doublet. 
The left panel represents the energy levels when considering the Coulomb interactions and the octahedral CEF under the conditions in Eq.~(\ref{eq:high-spin conditions}); the high-spin state with 12-fold degeneracy is split off from other higher-energy levels. The right panel shows the splitting of the 12-fold levels by the SOC in Eq.~(\ref{eq:H_SOC}). The lowest-energy levels are twofold degenerate, which comprise the Kramers doublet described by Eqs.~(\ref{eq:+}) and (\ref{eq:-}). 
}
\end{figure}

In the atomic state with the high-spin $d^7$ configuration, $\mathcal{H}_{\mathrm{SOC}}$ splits the 12-fold $t_{2g}^5e_{g}^{2}({}^3A_{2g}){}^{4}T_{1g}$ levels into three manifolds, as schematically shown in Fig.~\ref{fig:split}. 
The lowest-energy one is doublet, described by 
\begin{align}
\label{eq:+}
\ket{+}
&=\frac{i}{2}\Ket{\frac{3}{2},
x}+\frac{1}{2}\Ket{\frac{3}{2},y}-\frac{i}{\sqrt{3}}\Ket{\frac{1}{2},z}\nonumber\\
&-\frac{i}{2\sqrt{3}}\Ket{-\frac{1}{2},x}+\frac{1}{2\sqrt{3}}\Ket{-\frac{1}{2},y},\\
\label{eq:-}
\ket{-} &=-\frac{i}{2}\Ket{-\frac{3}{2},x}+\frac{1}{2}\Ket{-\frac{3}{2},y}-\frac{i}{\sqrt{3}}\Ket{-\frac{1}{2},z}\nonumber\\
&+\frac{i}{2\sqrt{3}}\Ket{\frac{1}{2},x}+\frac{1}{2\sqrt{3}}\Ket{\frac{1}{2},y}.
\end{align}

It is worth noting that the two states in Eqs.~(\ref{eq:+}) and (\ref{eq:-}) comprise a time-reversal Kramers pair: 
\begin{align}
\Theta \ket{+}=\ket{-}, \quad \Theta \ket{-}=-\ket{+}, 
\label{eq:Kramers}
\end{align}  
where $\Theta$ is the time-reversal operator satisfying
\begin{align}
\Theta \chi \ket{0}=\chi^*\ket{0}, \quad \Theta c^{\dagger}_{\alpha \sigma}\Theta^{-1}=(-1)^{\frac{1-\sigma}{2}}c^{\dagger}_{\alpha -\sigma},
\label{eq:time}
\end{align}
where $\chi \in \mathbb{C}$ and $\sigma=\pm1$. 
When we define $\tilde{J}^\mu= \tilde{L}^\mu +\tilde{S}^\mu$  ($\mu=x,y,z$), the  Kramers doublet $\ket\pm$ satisfy the following relations:
\begin{align}
\tilde{J}^z\ket{\pm}=\pm \frac{1}{2}\ket{\pm},
\quad
\tilde{J}^\pm\ket{\mp}=\ket{\pm},
\quad
\tilde{J}^\pm\ket{\pm}=0,
\end{align}
where $\tilde{J}^\pm = \tilde{J}^x \pm i\tilde{J}^y$.
These relations allow us to regard ${\bf \tilde{J}}$ as a fictitious angular momentum operator acting on the pseudo spin-half space described by $\ket\pm$.
The situation is similar to the so-called $J_{\mathrm{eff}}=1/2$ states discussed for the low-spin $d^5$ configuration~\cite{PhysRevLett.102.017205}, but the pseudo spin-half states $|\pm\rangle$ for the current $d^7$ situation are different from those for the $d^5$ state.

Thus, we end up with the low-energy Kramers doublet $|\pm\rangle$ for the high-spin $d^7$ state, which can be treated as the pseudo spin-half degrees of freedom. 
This is achieved by considering the situation with $U,J,\Delta \gg \lambda$ in the atomic limit of $\mathcal{H}_{\mathrm{hop}}=0$ under the conditions in Eq.~(\ref{eq:high-spin conditions}). 
In the next subsection, we show how to derive an effective low-energy Hamiltonian by introducing $\mathcal{H}_{\mathrm{hop}}$ as a perturbation.

\subsection{Second-order perturbation in terms of electron hopping} 
\label{subsec:second}

\begin{figure}[htp]
\centering
\includegraphics[width=\columnwidth,clip]{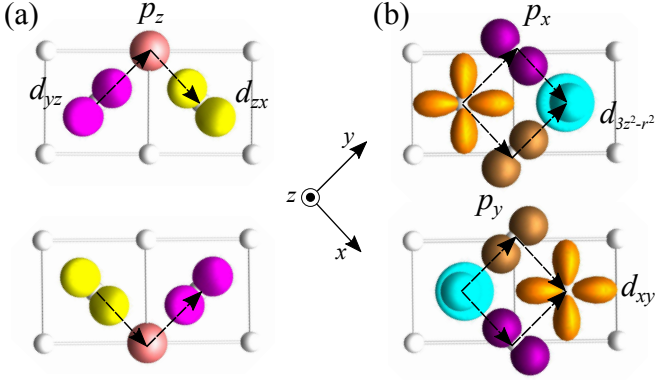}
\caption{\label{fig:hop} 
Schematic pictures for hopping processes via ligand ions taken into account in the Hamiltonian in Eq.~(\ref{eq:H^z_hop}): (a) $t_{2g}$-$t_{2g}$ hopping in the first term and (b) $t_{2g}$-$e_{g}$ hopping in the second term.  
The squares represent two edge-sharing octahedra viewed from the $z$ direction (the apical ligands are omitted) composed of the ligand ions (white spheres). 
The colored objects represent the $d$ and $p$ orbitals as indicated in the figures. 
}
\end{figure}

For the hopping term $\mathcal{H}_{\mathrm{hop}}$ in Eq.~(\ref{eq:H_multi}), we take into account the electron transfers between neighboring $M$ cations mediated by the ligands shared by the neighboring $MX_6$ octahedra. 
The explicit form depends on the $M$-$M$ bond direction because of the spatial anisotropy of $d$ orbitals as well as $p$ orbitals in the ligands. 
The situation is similar to the previous study for the low-spin $d^5$ state~\cite{PhysRevLett.102.017205}. 
For instance, for a bond lying on the $xy$ plane as shown in Fig.~\ref{fig:hop}, which we call the $z$ bond, the electron transfers have nonzero values between $d_{yz}$ and $d_{zx}$ orbitals via $p_z$ orbitals [Fig.~\ref{fig:hop}(a)] or $d_{xy}$ and $d_{3z^2-r^2}$ orbitals via $p_x$ and $p_y$ orbitals [Fig.~\ref{fig:hop}(b)]: 
all the other combinations vanish from the symmetry. 
Thus, the hopping term for the $z$ bond is given in the form
\begin{align}
\mathcal{H}^{(z)}_{\mathrm{hop}}=
\sum_{\langle i,j\rangle_{z},\sigma}&\left[t_{\xi \eta}(c^{\dagger}_{i\xi \sigma}c_{j\eta \sigma}+c^{\dagger}_{i\eta \sigma}c_{j\xi \sigma})\right.\nonumber\\
&\left.+
t_{\zeta u}(c^{\dagger}_{i\zeta \sigma}c_{ju\sigma}+c^{\dagger}_{iu \sigma}c_{j\zeta \sigma})+{\mathrm{h. c.}}\right].
\label{eq:H^z_hop}
\end{align}
Here, $t_{\xi\eta}$ and $t_{\zeta u}$ are represented in terms of the $p$-$d$ hopping integrals as 
\begin{align}
  t_{\xi\eta}=\frac{(pd\pi)^2}{\Delta_{pd}}, \quad 
  t_{\zeta u}=-\frac{(pd\pi)(pd\sigma)}{\Delta_{pd}},
\end{align}
where $(pd\pi)$ and $(pd\sigma)$ are the Slater-Koster parameters~\cite{PhysRev.94.1498} and $\Delta_{pd}$ describes the energy difference between $p$ and $d$ levels (we assume $\Delta \ll \Delta_{pd}$).
For other bond directions, the hopping term is obtained by cyclic permutations of $x$, $y$, and $z$, and $\mathcal{H}_{\mathrm{hop}}$ is given by the sum  
\begin{align}
\mathcal{H}_{\mathrm{hop}} = \sum_\mu \mathcal{H}^{(\mu)}_{\mathrm{hop}}.
\label{eq:H_hop}
\end{align}

Taking the Kramers doublet $\ket\pm$ as the ground states of the Hamiltonian in Eq.~(\ref{eq:H_multi}) in the atomic limit at each site, we perform the second-order perturbation in terms of electron hopping given by Eq.~(\ref{eq:H_hop}). 
Following the standard perturbation theory, the effective Hamiltonian for a $z$ bond connecting sites $i$ and $j$  is in general described as
\begin{equation}
h_{ij}^{(z)}=\sum_{a,b,c,d=\pm}\sum_n\frac{\bra{a,b}\mathcal{H}^{(z)}_{{\mathrm{hop}}}\ket{n}\bra{n}\mathcal{H}^{(z)}_{{\mathrm{hop}}}\ket{c,d}}{E_0-E_n}\ket{a,b}\bra{c,d}.
\label{eq:H_eff}
\end{equation}
Here, $|a,b\rangle$ and $|c,d\rangle$ describe two-site states within the ground state manifold composed of the Kramers doublet, whose energy is given by $E_0=2(21U-49J+2\Delta)$; 
the sum of $n$ runs over the intermediate states in the perturbation processes, and $E_n$ denotes the energy of the intermediate state. 

The intermediate states $\ket{n}$ in Eq.~(\ref{eq:H_eff}) are $\ket{d^8,d^6}$ (and $\ket{d^6,d^8}$) electron configurations generated by an electron hopping from the $\ket{d^7,d^7}$ ground states. 
For instance, by the $t_{2g}$-$t_{2g}$ hopping in the first term in Eq.~(\ref{eq:H^z_hop}), we obtain $\ket{t_{2g}^6e_{g}^{2}({}^3A_{2g}),t_{2g}^4e_{g}^{2}({}^3A_{2g})}$.
In this case, the $e_g$ electron configurations at both sites are unchanged by the electron transfers. 
On the other hand, for the $t_{2g}$-$e_g$ hopping in the second term in Eq.~(\ref{eq:H^z_hop}), we obtain $\ket{t_{2g}^5e_{g}^{3},t_{2g}^4e_{g}^{2}({}^3A_{2g})}$ or $\ket{t_{2g}^6e_{g}^{2}({}^3A_{2g}),t_{2g}^5e_{g}}$. 
In this case, the  $e_g$ electron configuration at one of the two sites is changed, while the $t_{2g}$ electron configuration at the same site remains unchanged. 
In the calculations, we neglect the energy splitting of the intermediate states by the SOC for simplicity, and obtain the effective Hamiltonian by neglecting the contributions of $\mathcal{O}(\lambda/U, \lambda/J, \lambda/\Delta)$.

We note that the intermediate states $\ket{n}$, which are the eigenstates of the unperturbed Hamiltonian $\mathcal{H}_{\mathrm{CEF}}+\mathcal{H}_{\mathrm{int}}+\mathcal{H}_{\mathrm{SOC}}$, are superpositions of different $\ket{d^8,d^6}$ configurations in general. 
This is because $\mathcal{H}_{\mathrm{int}}$ can have nonzero matrix elements between the different configurations; 
for instance, $\bra{t_{2g}^5e_g^1\, {}^3T_{1g}}\mathcal{H}_{\mathrm{int}}\ket{t_{2g}^3({}^3T_{2g})e_g^3\, {}^3T_{1g}}=\bra{t_{2g}^3({}^3T_{2g})e_g^3\, {}^3T_{1g}}\mathcal{H}_{\mathrm{int}}\ket{t_{2g}^5e_g^1\, {}^3T_{1g}}=\sqrt{2}J$.  
Such off-diagonal components of $\mathcal{H}_\mathrm{int}$ are called the configuration interactions. 
We take into account the configuration interactions in the following calculations [except in Eqs.~(\ref{eq:d5_K}) and (\ref{eq:d5_H})].
Note that the configuration interaction always vanishes for the $t_{2g}$-$t_{2g}$ hoppings.

\section{Result}
\label{sec:result}

In this section, we present the low-energy effective Hamiltonian for the pseudo spin-half degree of freedom for the high-spin $d^7$ system obtained by the perturbation theory in the previous section.
We also derive the effective Hamiltonian for the low-spin $d^5$ case in the similar perturbation scheme. 
In both cases, we obtain the Kitaev-Heisenberg Hamiltonian~\cite{PhysRevLett.105.027204,PhysRevLett.110.097204}:
\begin{equation}
    \mathcal{H}_{\mathrm{eff}} = 
    \sum_{\langle i,j \rangle_x } h_{ij}^{(x)}
    +\sum_{\langle i,j \rangle_y } h_{ij}^{(y)}
    +\sum_{\langle i,j \rangle_z } h_{ij}^{(z)},
    \label{eq:d7model}
  \end{equation}
  where
  \begin{equation}
    h_{ij}^{(\mu)} = J_{ K} \tilde{J}_i^\mu \tilde{J}_j^\mu +J_{ H}\tilde{\bf J}_i\cdot \tilde{\bf J}_j.
    \label{eq:h_ij}
\end{equation}
The first term in Eq.~(\ref{eq:h_ij}) describes the Kitaev-type anisotropic interaction as in Eq.~(\ref{eq:Kitaev_Hamiltonian}) (note that the sign is inverted), and the second term represents the isotropic Heisenberg interaction.
The explicit forms of the coupling constants $J_{ K}$ and $J_{ H}$ are shown in the following subsections.

We note that the ground-state phase diagram of the Kitaev-Heisenberg model in Eq.~(\ref{eq:d7model}) was studied using the exact diagonalization of a 24-site cluster~\cite{PhysRevLett.105.027204,PhysRevLett.110.097204}. Several phases, including the Kitaev QSL, were predicted depending on the values of $J_{ K}$ and $J_{H}$.
By referring to the previous results, we present the ground-state phase diagrams of the effective models for the $d^7$ and $d^5$ cases.

\subsection{High-spin $d^7$ case}
\label{subsec:effd7}

In the high-spin $d^7$ case, we obtain the coupling constants for the Kitaev and Heisenberg terms as
\begin{widetext}
\begin{align}
\label{eq:d7_J_K}
J_{ K}&=\left( -\frac{1}{3}\frac{1}{U-3J}+\frac{73}{243}\frac{1}{U+J}-\frac{4}{243}\frac{1}{U+4J}\right) t_{\xi \eta}^2\nonumber\\
&+\left(-\frac{5}{27}\frac{1}{U-3J+\Delta}+\frac{5}{243}\frac{1}{U+J+\Delta}-\frac{20}{243}\frac{1}{U+4J+\Delta}-\frac{20}{81}\frac{U+\Delta+3J}{U^2-\Delta^2+5JU-J\Delta+4J^2}\right)t_{\zeta u}^2,\\
J_{ H}&=
\left(-\frac{7}{18}\frac{1}{U-3J}+\frac{29}{54}\frac{1}{U+J}+\frac{2}{27}\frac{1}{U+4J}\right)t_{\xi \eta}^2\nonumber\\
&+\left(-\frac{5}{9}\frac{1}{U-3J+\Delta}+\frac{215}{243}\frac{1}{U+J+\Delta}+\frac{40}{243}\frac{1}{U+4J+\Delta}+\frac{40}{81}\frac{U+\Delta+3J}{U^2-\Delta^2+5JU-J\Delta+4J^2}\right)t_{\zeta u}^2,
\label{eq:d7_J_H}
\end{align}
\end{widetext} 
respectively. 
The first terms proportional to $t_{\xi \eta}^2$ in both Eqs.~(\ref{eq:d7_J_K}) and (\ref{eq:d7_J_H}) originate from the hopping processes between $t_{2g}$ orbitals, and the second ones proportional to $t_{\zeta u}^2$ are from the hopping processes between $t_{2g}$ and $e_g$ orbitals. 
We find that the Kitaev interaction in Eq.~(\ref{eq:d7_J_K}) is always ferromagnetic, $J_{ K} < 0$, for the parameter range considered here,
while the Heisenberg interaction $J_H$ in Eq.~(\ref{eq:d7_J_H}) can be both ferromagnetic and antiferromagnetic. 
Note that the SOC $\lambda$ does not appear in the expressions as we omit the energy splitting by $\lambda$ in the intermediate states in the perturbations.

\begin{figure}[htp]
\centering
\includegraphics[width=\columnwidth,clip]{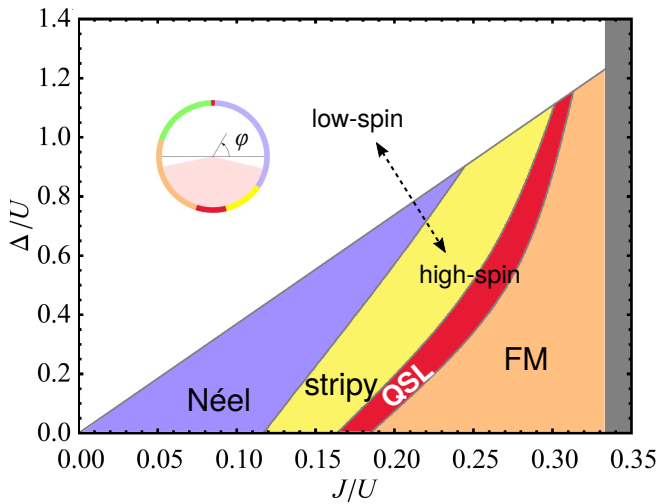}
\caption{\label{fig:d7} 
The phase diagram for the high-spin $d^7$ model. 
We take $t_{\zeta u}/t_{\xi \eta}=2$. 
The white region indicates the parameter range where the system is in the low-spin state in the atomic limit. 
The gray region represents the parameter range out of the conditions in Eq.~(\ref{eq:high-spin conditions}). 
The inset shows the phase diagram obtained in Ref.~\cite{PhysRevLett.110.097204}: 
the colors correspond to those in the main panel. 
The green, orange, yellow, purple, and red regions correspond to the zigzag, ferromagnetic (FM), stripy, N\'eel, and QSL phases. 
The shaded range of $\varphi$ in the inset represents the physically-reasonable parameter range for the $d^7$ model. 
The four phases in this range appear in the phase diagram in the main panel. 
}
\end{figure}

Figure~\ref{fig:d7} shows the ground-state phase diagram  for the Kitaev-Heisenberg model for the high-spin $d^7$ case given by Eq.~(\ref{eq:d7model}) with Eqs.~(\ref{eq:d7_J_K}) and (\ref{eq:d7_J_H}). 
The result is obtained by referring to the previous result in Ref.~\cite{PhysRevLett.110.097204}: 
for convenience, we present the previous result in the inset where the parameter $\varphi$ is defined as 
\begin{align}
\varphi=\mbox{sgn}(J_{ K})\arccos \frac{J_{ H}}{\sqrt{\big(\frac{J_{ K}}{2}\big)^2+J_{ H}^2}}.
\end{align}
Here, we set $t_{\zeta u}/t_{\xi \eta}=2$ by using the relation $(pd\sigma)/(pd\pi) \approx -2$~\cite{harrison2012electronic}. 
The shaded range of $\varphi$ in the inset indicates the physically-reasonable parameter range for our $d^7$ model, which is limited by the two conditions in Eq.~(\ref{eq:high-spin conditions}). 
In the main panel of Fig.~\ref{fig:d7}, the gray region represents $J/U>1/3$ and the white region $\Delta/J >\sqrt{22}-1$.

As shown in Fig.~\ref{fig:d7}, the Kitaev-Heisenberg model for the high-spin $d^7$ state exhibits four different phases: 
three magnetically ordered phases and a QSL phase. 
In the small $\Delta$ and $J$ region, the system shows the N\'eel-type antiferromagnetic order, while it is replaced by the stripy order by increasing $J$. 
With a further increase of $J$, the system turns into the ferromagnetic state, but before entering it, there is a window for the QSL between the stripe and ferromagnetic phases. 
Thus, the high-spin $d^7$ case provides another chance to realize the QSL, in addition to the low-spin $d^5$ case studied so far.

\subsection{Low-spin $d^5$ case}
\label{subsec:d5}

For comparison, we here derive the effective Hamiltonian for the low-spin $d^5$ case by using the similar framework of the perturbation. 
We note that such a Hamiltonian was already derived in the previous study~\cite{PhysRevLett.102.017205}, but the full form was not shown explicitly in the literature:
the effective Hamiltonian was shown for the case by considering only the first term of Eq.~(\ref{eq:H^z_hop}) in the perturbation. 
Meanwhile, the effective Hamiltonian arising from the second term of Eq.~(\ref{eq:H^z_hop}) was derived in the limit of $J/U \ll 1$ in Ref.~\cite{PhysRevLett.110.097204}, but the contributions from the first term were omitted. 
We here present the full form of the effective Hamiltonian by including both the first and second terms in Eq.~(\ref{eq:H^z_hop}).
By neglecting the configuration interactions  for simplicity, we obtain the coupling constants as 
\begin{widetext}
\begin{align}
J_{ K}&=\left(-\frac{4}{3}\frac{1}{U-3J}+\frac{4}{3}\frac{1}{U-J}\right)t_{\xi \eta}^2
+\left(-\frac{4}{27}\frac{1}{U+\Delta+J}-\frac{2}{27}\frac{1}{U+\Delta-2 J}-\frac{4}{27}\frac{1}{U+\Delta-4 J}+\frac{2}{9}\frac{1}{U+\Delta-6 J}+\frac{4}{27}\frac{1}{U+\Delta-J}\right)t^2_{\zeta u},
\label{eq:d5_K}
\\
J_{ H}
&=\left(\frac{2}{27}\frac{1}{U+\Delta+J}+\frac{4}{27}\frac{1}{U+\Delta-2 J}-\frac{4}{27}\frac{1}{U+\Delta-4 J}-\frac{4}{27}\frac{1}{U+\Delta-J}\right)t^2_{\zeta u}.
\label{eq:d5_H}
\end{align}
\end{widetext}
The first term proportional to $t_{\xi\eta}^2$ in Eq.~(\ref{eq:d5_K}) is equivalent to the expression shown in Ref.~\cite{PhysRevLett.102.017205}. 
In addition, the second term of Eqs.~(\ref{eq:d5_K}) and (\ref{eq:d5_H}) are both consistent with the expressions in Ref.~\cite{PhysRevLett.110.097204} in the limit of $J/U \ll 1$.
When we take $t_{\zeta u}/t_{\xi \eta}=2$ as in the $d^7$ case above, we find that the Kitaev interaction can be both ferromagnetic and antiferromagnetic, while the Heisenberg interaction is always ferromagnetic, $J_{ H} < 0$.

As in the $d^7$ case, we obtain the ground-state phase diagram for the effective Kitaev-Heisenberg model for the $d^5$ case with Eqs.~(\ref{eq:d5_K}) and (\ref{eq:d5_H}). 
In the calculations,  we numerically estimate $J_{ K}$ and $J_{ H}$ by taking into account the configuration interactions in the intermediate states. 

\begin{figure}[htp]
\centering
\includegraphics[width=\columnwidth,clip]{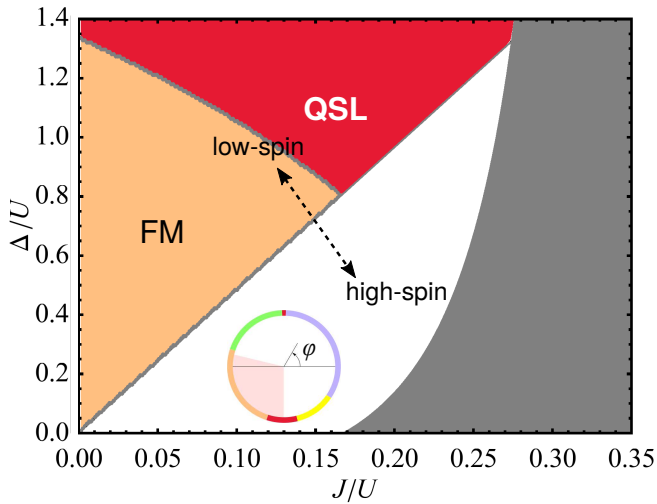}
\caption{\label{fig:d5} The phase diagram for the low-spin $d^5$ model. We take $t_{\zeta u}/t_{\xi \eta}=2$. 
The white region indicates the parameter range where the system is in the high-spin state in the atomic limit. The gray region represents the parameter range where 
$E_n < E_0$ for one of the intermediate states $\ket{n}$. 
The inset shows the phase diagram obtained in Ref.~\cite{PhysRevLett.110.097204}: the colors correspond to those in the main panel. 
The notations are common to those in Fig.~\ref{fig:d7}.
}
\end{figure}

Figure~\ref{fig:d5} shows the ground-state phase diagram  for the Kitaev-Heisenberg model for the low-spin $d^5$ case given by Eq.~(\ref{eq:d7model}) with Eqs.~(\ref{eq:d5_K}) and (\ref{eq:d5_H}). 
The result is obtained by referring to the previous result in Ref.~\cite{PhysRevLett.110.097204} as in the $d^7$ case in the previous subsection.
The shaded range of $\varphi$ in the inset indicates the physically-reasonable parameter range for our $d^5$ model, which is limited by two conditions: 
$ E_n-E_0<0$ for an intermediate state $\ket{n}$ (the gray region in Fig.~\ref{fig:d5}) and $\Delta/J <\sqrt{70/3}$ for assuring the low-spin state (the white region in Fig.~\ref{fig:d5}).
As shown in Fig.~\ref{fig:d5}, the Kitaev-Heisenberg model for the low-spin $d^5$ state exhibits only two phases in the physically-reasonable parameter region, in contrast to the $d^7$ case: 
a ferromagnetically ordered phase and a QSL phase.

\section{Discussion}
\label{dis}

Let us discuss our results for the high-spin $d^7$ case in comparison with those for the low-spin $d^5$ case. 
Comparing the phase diagrams in Figs.~\ref{fig:d7} and \ref{fig:d5}, we note that the phase diagram for the $d^7$ case is richer than that for the $d^5$ case: the $d^7$ case includes four phases in addition to the QSL, while the $d^5$ case includes only two. 
In addition, the former includes both ferromagnetic and antiferromagnetic phases. 
This is due to the fact that in the $d^7$ model the Heisenberg coupling $J_H$ changes the sign by tuning $J/U$ and $\Delta/U$ [Eq.~(\ref{eq:d7_J_H})]. 

\begin{figure}[htp]
\centering
\includegraphics[width=0.9\columnwidth,clip]{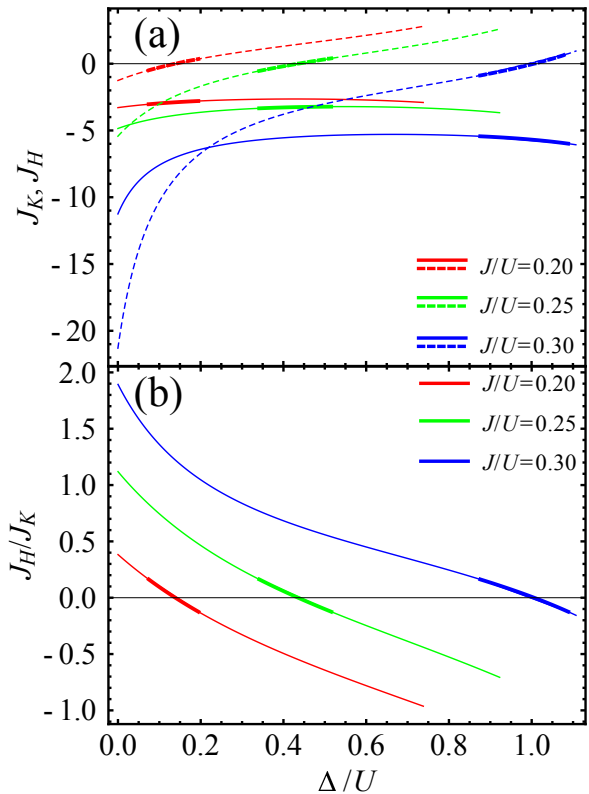}
\caption{\label{fig:deltad7H}  
(a) $J_K$ and $J_H$ in units of $t_{\xi \eta}^2/U$ [Eqs.~(\ref{eq:d7_J_K}) and (\ref{eq:d7_J_H})] for several values of $J/U$ in the high-spin $d^7$ case. 
The solid and dashed lines represent $J_K$ and $J_H$, respectively. 
The thick parts indicate the QSL regions.
(b) The ratio $J_H/J_K$. 
We take $t_{\zeta u}/t_{\xi \eta}=2$. 
}
\end{figure}

In order to look closer on how the interactions change, we plot the values of $J_K$ and $J_H$ (in units of $t^2_{\xi\eta}/U$) for several $J/U$ in the $d^7$ case in Fig.~\ref{fig:deltad7H}(a). 
As mentioned in Sec.~\ref{subsec:effd7}, the Kitaev interaction $J_K$ is always negative (ferromagnetic), while the Heisenberg interaction $J_H$ changes its sign from negative (ferromagnetic) to positive (antiferromagnetic) as $\Delta/U$ increases. 
In the QSL region (thick lines in the plot), $J_H$ almost vanishes and $J_K$ becomes dominant. 
We plot the ratio between $J_H$ and $J_K$ in Fig.~\ref{fig:deltad7H}(b), which also changes the sign and is minimized in the QSL region.

\begin{figure}[htp]
\centering
\includegraphics[width=0.9\columnwidth,clip]{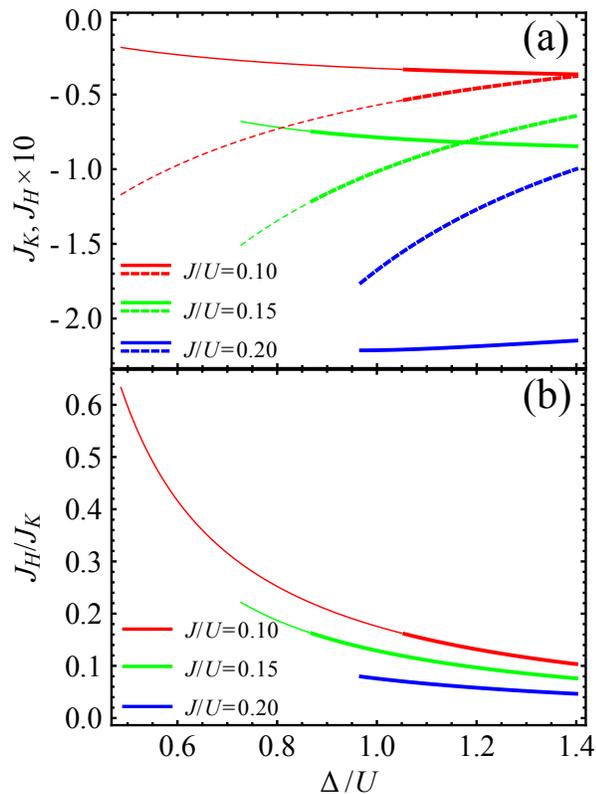}
\caption{\label{fig:deltad5H}
(a) $J_K$ and $J_H$ in units of $t_{\xi \eta}^2/U$ [Eqs.~(\ref{eq:d5_K}) and (\ref{eq:d5_H})] for several values of $J/U$ in the low-spin $d^5$ case; 
$J_H$ is multiplied by $10$ for visibility. 
The solid and dashed lines represent $J_K$ and $J_H$, respectively. 
The thick parts indicate the QSL regions. 
(b) The ratio $J_H/J_K$. 
We take $t_{\zeta u}/t_{\xi \eta}=2$. 
}
\end{figure}

In the $d^5$ case, $J_K$ can be both positive and negative, but the region for $J_K>0$ is limited to small $J/U$ and $\Delta/U$. 
Meanwhile, $J_H$ is always negative and never vanishes, in contrast to the $d^7$ case. 
We plot the typical values of $J_K$ and $J_H$ in Fig.~\ref{fig:deltad5H}(a). 
The ratio $J_H/J_K$ is plotted in Fig.~\ref{fig:deltad5H}(b), which is positive in this range in contrast to the $d^7$ case in Fig.~\ref{fig:deltad7H}(b). 

The comparison above indicates that the $d^7$ case has the following advantages for realizing the Kitaev QSL compared to the $d^5$ case. 
First, the Heisenberg interaction $J_H$, which perturbs the Kitaev QSL, can be minimized and even eliminated by tuning the parameters in the $d^7$ case. 
In the $d^5$ case, $J_H$ can be small but never vanishes. 
Another advantage is that the bare energy scale of $J_K$ can be larger for the $d^7$ case, as shown in Figs.~\ref{fig:deltad7H}(a) and \ref{fig:deltad5H}(a), assuming that the energy unit $t_{\xi\eta}^2/U$ is roughly the same in both cases. 
The magnitude of $J_K$ determines the temperature and energy scales, where a salient feature of the Kitaev QSL, the fractionalization of quantum spins into Majorana fermions, sets in~\cite{PhysRevB.92.115122}. 
Hence, the larger $J_K$ may makes the high-spin $d^7$ systems more suitable to observe the spin fractionalization in the Kitaev QSL. 

\section{Summary}
\label{sum}

In summary, we have proposed a platform for the Kitaev QSL by considering the high-spin $d^7$ electron configuration. 
The atomic ground state of the $d^7$ case comprises the spin-orbital entangled Kramers pair, which is different from the one considered for the low-spin $d^5$ case in the previous studies. 
Using the perturbation in terms of $d$-$p$-$d$ hoppings for edge-sharing octahedra, we showed that the effective spin model in the strong coupling limit gives rise to the Kitaev-Heisenberg model. 
Referring to the previous study~\cite{PhysRevLett.110.097204}, we constructed the ground-state phase diagram for this $d^7$ model as a function of the Hund's-rule coupling and the crystalline electric field splitting.  
We found that the model exhibits the QSL phase in the physically reasonable parameter range, in addition to three magnetically ordered phases. 
Thus, our results extend the candidates for the Kitaev QSL, beyond the low-spin $d^5$ compounds studied thus far.

For comparison, we have also studied the ground-state phase diagram for the effective model for the low-spin $d^5$ electron configuration derived by a similar perturbation. 
We found that the $d^7$ case exhibits richer phases than the $d^5$ case. 
This is due to different dependences of the Kitaev and Heisenberg couplings, $J_K$ and $J_H$, respectively, on the Coulomb interactions and the crystalline electric field splitting.
In the $d^7$ case, $J_K$ is always ferromagnetic, while $J_H$ can be either ferromagnetic or antiferromagnetic. 
Meanwhile, in the $d^5$ case, $J_K$ can be either ferromagnetic or antiferromagnetic, while $J_H$ is always ferromagnetic. 
Thus, the Heisenberg interaction $J_H$, which perturbs the Kitaev QSL, can be minimized even to zero in the $d^7$ case. 
In addition, we found that the bare energy scale of the Kitaev interaction $J_K$ can be larger in the $d^7$ case compared to the $d^5$ case. 
These are advantages of the $d^7$ case for realizing the Kitaev QSL and to experimentally observe the salient feature, the fractionalization of spins into Majorana fermions. 

Candidate materials for the high-spin $d^7$ case would be the compounds including Co$^{2+}$ and Ni$^{3+}$ cations. 
In the $3d$ electron systems, the spin-orbit coupling is weaker compared to the $4d$ and $5d$ cases, but the strong electron correlations are preferable to stabilize the high-spin configuration. 
Our results will stimulate such material search for extending the physics of Kitaev QSLs.

{\it Note added.} During writing the manuscript, we noticed that a similar scenario has been proposed independently by Liu and Khaliullin~\cite{liu2017}.

\begin{acknowledgments}
The authors thank J. Yoshitake for constructive suggestions. 
R. S. thanks A. Matsugatani, P. A. Mishchenko and M. G. Yamada for fruitful discussions. 
This work was supported by JSPS KAKENHI Grants No.~JP15K13533, No.~JP16K17747, and No.~JP16H02206. 
\end{acknowledgments}
\bibliography{d7}

\begin{thebibliography}{30}%
\makeatletter
\providecommand \@ifxundefined [1]{%
 \@ifx{#1\undefined}
}%
\providecommand \@ifnum [1]{%
 \ifnum #1\expandafter \@firstoftwo
 \else \expandafter \@secondoftwo
 \fi
}%
\providecommand \@ifx [1]{%
 \ifx #1\expandafter \@firstoftwo
 \else \expandafter \@secondoftwo
 \fi
}%
\providecommand \natexlab [1]{#1}%
\providecommand \enquote  [1]{``#1''}%
\providecommand \bibnamefont  [1]{#1}%
\providecommand \bibfnamefont [1]{#1}%
\providecommand \citenamefont [1]{#1}%
\providecommand \href@noop [0]{\@secondoftwo}%
\providecommand \href [0]{\begingroup \@sanitize@url \@href}%
\providecommand \@href[1]{\@@startlink{#1}\@@href}%
\providecommand \@@href[1]{\endgroup#1\@@endlink}%
\providecommand \@sanitize@url [0]{\catcode `\\12\catcode `\$12\catcode
  `\&12\catcode `\#12\catcode `\^12\catcode `\_12\catcode `\%12\relax}%
\providecommand \@@startlink[1]{}%
\providecommand \@@endlink[0]{}%
\providecommand \url  [0]{\begingroup\@sanitize@url \@url }%
\providecommand \@url [1]{\endgroup\@href {#1}{\urlprefix }}%
\providecommand \urlprefix  [0]{URL }%
\providecommand \Eprint [0]{\href }%
\providecommand \doibase [0]{http://dx.doi.org/}%
\providecommand \selectlanguage [0]{\@gobble}%
\providecommand \bibinfo  [0]{\@secondoftwo}%
\providecommand \bibfield  [0]{\@secondoftwo}%
\providecommand \translation [1]{[#1]}%
\providecommand \BibitemOpen [0]{}%
\providecommand \bibitemStop [0]{}%
\providecommand \bibitemNoStop [0]{.\EOS\space}%
\providecommand \EOS [0]{\spacefactor3000\relax}%
\providecommand \BibitemShut  [1]{\csname bibitem#1\endcsname}%
\let\auto@bib@innerbib\@empty
\bibitem [{\citenamefont {Anderson}(1973)}]{ANDERSON1973153}%
  \BibitemOpen
  \bibfield  {author} {\bibinfo {author} {\bibfnamefont {P.~W.}\ \bibnamefont
  {Anderson}},\ }\bibfield  {title} {\enquote {\bibinfo {title} {Resonating
  valence bonds: a new kind of insulator?{}},}\ }\href {\doibase
  http://dx.doi.org/10.1016/0025-5408(73)90167-0} {\bibfield  {journal}
  {\bibinfo  {journal} {Mater. Res. Bull.}\ }\textbf {\bibinfo {volume} {8}},\
  \bibinfo {pages} {153} (\bibinfo {year} {1973})}\BibitemShut {NoStop}%
\bibitem [{\citenamefont {Balents}(2010)}]{balents2010spin}%
  \BibitemOpen
  \bibfield  {author} {\bibinfo {author} {\bibfnamefont {L.}~\bibnamefont
  {Balents}},\ }\bibfield  {title} {\enquote {\bibinfo {title} {Spin {L}iquids
  in {F}rustrated {M}agnets},}\ }\href@noop {} {\bibfield  {journal} {\bibinfo
  {journal} {Nature}\ }\textbf {\bibinfo {volume} {464}},\ \bibinfo {pages}
  {199} (\bibinfo {year} {2010})}\BibitemShut {NoStop}%
\bibitem [{\citenamefont {Savary}\ and\ \citenamefont
  {Balents}(2017)}]{0034-4885-80-1-016502}%
  \BibitemOpen
  \bibfield  {author} {\bibinfo {author} {\bibfnamefont {L.}~\bibnamefont
  {Savary}}\ and\ \bibinfo {author} {\bibfnamefont {L.}~\bibnamefont
  {Balents}},\ }\bibfield  {title} {\enquote {\bibinfo {title} {Quantum spin
  liquids: a review},}\ }\href
  {http://stacks.iop.org/0034-4885/80/i=1/a=016502} {\bibfield  {journal}
  {\bibinfo  {journal} {Rep. Prog. Phys}\ }\textbf {\bibinfo {volume} {80}},\
  \bibinfo {pages} {016502} (\bibinfo {year} {2017})}\BibitemShut {NoStop}%
\bibitem [{\citenamefont {Zhou}\ \emph {et~al.}(2017)\citenamefont {Zhou},
  \citenamefont {Kanoda},\ and\ \citenamefont {Ng}}]{RevModPhys.89.025003}%
  \BibitemOpen
  \bibfield  {author} {\bibinfo {author} {\bibfnamefont {Y.}~\bibnamefont
  {Zhou}}, \bibinfo {author} {\bibfnamefont {K.}~\bibnamefont {Kanoda}}, \ and\
  \bibinfo {author} {\bibfnamefont {T.-K.}\ \bibnamefont {Ng}},\ }\bibfield
  {title} {\enquote {\bibinfo {title} {Quantum {S}pin {L}iquid {S}tates},}\
  }\href {\doibase 10.1103/RevModPhys.89.025003} {\bibfield  {journal}
  {\bibinfo  {journal} {Rev. Mod. Phys.}\ }\textbf {\bibinfo {volume} {89}},\
  \bibinfo {pages} {025003} (\bibinfo {year} {2017})}\BibitemShut {NoStop}%
\bibitem [{\citenamefont {Sachdev}(1992)}]{sachdev1992kagome}%
  \BibitemOpen
  \bibfield  {author} {\bibinfo {author} {\bibfnamefont {S.}~\bibnamefont
  {Sachdev}},\ }\bibfield  {title} {\enquote {\bibinfo {title} {Kagom{\'e}- and
  triangular-lattice heisenberg antiferromagnets: Ordering from quantum
  fluctuations and quantum-disordered ground states with unconfined bosonic
  spinons},}\ }\href {\doibase 10.1103/PhysRevB.45.12377} {\bibfield  {journal}
  {\bibinfo  {journal} {Phys. Rev. B}\ }\textbf {\bibinfo {volume} {45}},\
  \bibinfo {pages} {12377} (\bibinfo {year} {1992})}\BibitemShut {NoStop}%
\bibitem [{\citenamefont {Hallberg}\ and\ \citenamefont
  {Balseiro}(1991)}]{PhysRevB.43.10289}%
  \BibitemOpen
  \bibfield  {author} {\bibinfo {author} {\bibfnamefont {K.}~\bibnamefont
  {Hallberg}}\ and\ \bibinfo {author} {\bibfnamefont {C.~A.}\ \bibnamefont
  {Balseiro}},\ }\bibfield  {title} {\enquote {\bibinfo {title} {Anyons in spin
  liquids},}\ }\href {\doibase 10.1103/PhysRevB.43.10289} {\bibfield  {journal}
  {\bibinfo  {journal} {Phys. Rev. B}\ }\textbf {\bibinfo {volume} {43}},\
  \bibinfo {pages} {10289} (\bibinfo {year} {1991})}\BibitemShut {NoStop}%
\bibitem [{\citenamefont {Kitaev}(2006)}]{Kitaev20062}%
  \BibitemOpen
  \bibfield  {author} {\bibinfo {author} {\bibfnamefont {A.}~\bibnamefont
  {Kitaev}},\ }\bibfield  {title} {\enquote {\bibinfo {title} {{A}nyons in an
  exactly solved model and beyond},}\ }\href {\doibase
  http://dx.doi.org/10.1016/j.aop.2005.10.005} {\bibfield  {journal} {\bibinfo
  {journal} {Ann. Phys. (N. Y.)}\ }\textbf {\bibinfo {volume} {321}},\ \bibinfo
  {pages} {2} (\bibinfo {year} {2006})}\BibitemShut {NoStop}%
\bibitem [{\citenamefont {Baskaran}\ \emph {et~al.}(2007)\citenamefont
  {Baskaran}, \citenamefont {Mandal},\ and\ \citenamefont
  {Shankar}}]{PhysRevLett.98.247201}%
  \BibitemOpen
  \bibfield  {author} {\bibinfo {author} {\bibfnamefont {G.}~\bibnamefont
  {Baskaran}}, \bibinfo {author} {\bibfnamefont {S.}~\bibnamefont {Mandal}}, \
  and\ \bibinfo {author} {\bibfnamefont {R.}~\bibnamefont {Shankar}},\
  }\bibfield  {title} {\enquote {\bibinfo {title} {Exact {R}esults for {S}pin
  {D}ynamics and {F}ractionalization in the {K}itaev {M}odel},}\ }\href
  {\doibase 10.1103/PhysRevLett.98.247201} {\bibfield  {journal} {\bibinfo
  {journal} {Phys. Rev. Lett.}\ }\textbf {\bibinfo {volume} {98}},\ \bibinfo
  {pages} {247201} (\bibinfo {year} {2007})}\BibitemShut {NoStop}%
\bibitem [{\citenamefont {Jackeli}\ and\ \citenamefont
  {Khaliullin}(2009)}]{PhysRevLett.102.017205}%
  \BibitemOpen
  \bibfield  {author} {\bibinfo {author} {\bibfnamefont {G.}~\bibnamefont
  {Jackeli}}\ and\ \bibinfo {author} {\bibfnamefont {G.}~\bibnamefont
  {Khaliullin}},\ }\bibfield  {title} {\enquote {\bibinfo {title} {Mott
  {I}nsulators in the {S}trong {S}pin-{O}rbit {C}oupling {L}imit: {F}rom
  {H}eisenberg to a {Q}uantum {Q}ompass and {K}itaev {M}odels},}\ }\href
  {\doibase 10.1103/PhysRevLett.102.017205} {\bibfield  {journal} {\bibinfo
  {journal} {Phys. Rev. Lett.}\ }\textbf {\bibinfo {volume} {102}},\ \bibinfo
  {pages} {017205} (\bibinfo {year} {2009})}\BibitemShut {NoStop}%
\bibitem [{\citenamefont {Khaliullin}(2005)}]{PTPS.160.155}%
  \BibitemOpen
  \bibfield  {author} {\bibinfo {author} {\bibfnamefont {G.}~\bibnamefont
  {Khaliullin}},\ }\bibfield  {title} {\enquote {\bibinfo {title} {Orbital
  {O}rder and {F}luctuations in {M}ott {I}nsulators},}\ }\href {\doibase
  10.1143/PTPS.160.155} {\bibfield  {journal} {\bibinfo  {journal} {Prog of
  Theor. Phys. Supplement}\ }\textbf {\bibinfo {volume} {160}},\ \bibinfo
  {pages} {155} (\bibinfo {year} {2005})}\BibitemShut {NoStop}%
\bibitem [{\citenamefont {Chaloupka}\ \emph {et~al.}(2010)\citenamefont
  {Chaloupka}, \citenamefont {Jackeli},\ and\ \citenamefont
  {Khaliullin}}]{PhysRevLett.105.027204}%
  \BibitemOpen
  \bibfield  {author} {\bibinfo {author} {\bibfnamefont {J.}~\bibnamefont
  {Chaloupka}}, \bibinfo {author} {\bibfnamefont {G.}~\bibnamefont {Jackeli}},
  \ and\ \bibinfo {author} {\bibfnamefont {G.}~\bibnamefont {Khaliullin}},\
  }\bibfield  {title} {\enquote {\bibinfo {title} {Kitaev-{H}eisenberg {M}odel
  on a {H}oneycomb {L}attice: {P}ossible {E}xotic {P}hases in {I}ridium
  {O}xides {A}$_{2}${I}r{O}$_{3}$},}\ }\href {\doibase
  10.1103/PhysRevLett.105.027204} {\bibfield  {journal} {\bibinfo  {journal}
  {Phys. Rev. Lett.}\ }\textbf {\bibinfo {volume} {105}},\ \bibinfo {pages}
  {027204} (\bibinfo {year} {2010})}\BibitemShut {NoStop}%
\bibitem [{\citenamefont {Chaloupka}\ \emph {et~al.}(2013)\citenamefont
  {Chaloupka}, \citenamefont {Jackeli},\ and\ \citenamefont
  {Khaliullin}}]{PhysRevLett.110.097204}%
  \BibitemOpen
  \bibfield  {author} {\bibinfo {author} {\bibfnamefont {J.}~\bibnamefont
  {Chaloupka}}, \bibinfo {author} {\bibfnamefont {G.}~\bibnamefont {Jackeli}},
  \ and\ \bibinfo {author} {\bibfnamefont {G.}~\bibnamefont {Khaliullin}},\
  }\bibfield  {title} {\enquote {\bibinfo {title} {{Z}igzag {M}agnetic {O}rder
  in the {I}ridium {O}xide {N}a$_{2}${I}r{O}$_{3}$},}\ }\href {\doibase
  10.1103/PhysRevLett.110.097204} {\bibfield  {journal} {\bibinfo  {journal}
  {Phys. Rev. Lett.}\ }\textbf {\bibinfo {volume} {110}},\ \bibinfo {pages}
  {097204} (\bibinfo {year} {2013})}\BibitemShut {NoStop}%
\bibitem [{\citenamefont {Rau}\ \emph {et~al.}(2014)\citenamefont {Rau},
  \citenamefont {Lee},\ and\ \citenamefont {Kee}}]{PhysRevLett.112.077204}%
  \BibitemOpen
  \bibfield  {author} {\bibinfo {author} {\bibfnamefont {J.~G.}\ \bibnamefont
  {Rau}}, \bibinfo {author} {\bibfnamefont {E.~K.-H.}\ \bibnamefont {Lee}}, \
  and\ \bibinfo {author} {\bibfnamefont {H.-Y.}\ \bibnamefont {Kee}},\
  }\bibfield  {title} {\enquote {\bibinfo {title} {Generic {S}pin {M}odel for
  the {H}oneycomb {I}ridates beyond the {K}itaev {L}imit},}\ }\href {\doibase
  10.1103/PhysRevLett.112.077204} {\bibfield  {journal} {\bibinfo  {journal}
  {Phys. Rev. Lett.}\ }\textbf {\bibinfo {volume} {112}},\ \bibinfo {pages}
  {077204} (\bibinfo {year} {2014})}\BibitemShut {NoStop}%
\bibitem [{\citenamefont {Sizyuk}\ \emph {et~al.}(2014)\citenamefont {Sizyuk},
  \citenamefont {Price}, \citenamefont {W\"olfle},\ and\ \citenamefont
  {Perkins}}]{PhysRevB.90.155126}%
  \BibitemOpen
  \bibfield  {author} {\bibinfo {author} {\bibfnamefont {Y.}~\bibnamefont
  {Sizyuk}}, \bibinfo {author} {\bibfnamefont {C.}~\bibnamefont {Price}},
  \bibinfo {author} {\bibfnamefont {P.}~\bibnamefont {W\"olfle}}, \ and\
  \bibinfo {author} {\bibfnamefont {N.~B.}\ \bibnamefont {Perkins}},\
  }\bibfield  {title} {\enquote {\bibinfo {title} {Importance of anisotropic
  exchange interactions in honeycomb iridates: minimal model for zigzag
  antiferromagnetic order in {N}a$_{2}${I}r{O}$_{3}$},}\ }\href {\doibase
  10.1103/PhysRevB.90.155126} {\bibfield  {journal} {\bibinfo  {journal} {Phys.
  Rev. B}\ }\textbf {\bibinfo {volume} {90}},\ \bibinfo {pages} {155126}
  (\bibinfo {year} {2014})}\BibitemShut {NoStop}%
\bibitem [{\citenamefont {Kim}\ \emph {et~al.}(2015)\citenamefont {Kim},
  \citenamefont {Vijay~Shankar}, \citenamefont {Catuneanu},\ and\ \citenamefont
  {Kee}}]{PhysRevB.91.241110}%
  \BibitemOpen
  \bibfield  {author} {\bibinfo {author} {\bibfnamefont {H.-S.}\ \bibnamefont
  {Kim}}, \bibinfo {author} {\bibfnamefont {V.}~\bibnamefont {Vijay~Shankar}},
  \bibinfo {author} {\bibfnamefont {A.}~\bibnamefont {Catuneanu}}, \ and\
  \bibinfo {author} {\bibfnamefont {H.-Y.}\ \bibnamefont {Kee}},\ }\bibfield
  {title} {\enquote {\bibinfo {title} {{K}itaev magnetism in honeycomb
  {R}u{C}l$_{3}$ with intermediate spin-orbit coupling},}\ }\href {\doibase
  10.1103/PhysRevB.91.241110} {\bibfield  {journal} {\bibinfo  {journal} {Phys.
  Rev. B}\ }\textbf {\bibinfo {volume} {91}},\ \bibinfo {pages} {241110}
  (\bibinfo {year} {2015})}\BibitemShut {NoStop}%
\bibitem [{\citenamefont {Chaloupka}\ and\ \citenamefont
  {Khaliullin}(2016)}]{PhysRevB.94.064435}%
  \BibitemOpen
  \bibfield  {author} {\bibinfo {author} {\bibfnamefont {J.}~\bibnamefont
  {Chaloupka}}\ and\ \bibinfo {author} {\bibfnamefont {G.}~\bibnamefont
  {Khaliullin}},\ }\bibfield  {title} {\enquote {\bibinfo {title} {Magnetic
  anisotropy in the {K}itaev model systems {N}a$_{2}${I}r{O}$_{3}$ and
  {R}u{C}l$_{3}$},}\ }\href {\doibase 10.1103/PhysRevB.94.064435} {\bibfield
  {journal} {\bibinfo  {journal} {Phys. Rev. B}\ }\textbf {\bibinfo {volume}
  {94}},\ \bibinfo {pages} {064435} (\bibinfo {year} {2016})}\BibitemShut
  {NoStop}%
\bibitem [{\citenamefont {Singh}\ and\ \citenamefont
  {Gegenwart}(2010)}]{PhysRevB.82.064412}%
  \BibitemOpen
  \bibfield  {author} {\bibinfo {author} {\bibfnamefont {Y.}~\bibnamefont
  {Singh}}\ and\ \bibinfo {author} {\bibfnamefont {P.}~\bibnamefont
  {Gegenwart}},\ }\bibfield  {title} {\enquote {\bibinfo {title}
  {Antiferromagnetic {M}ott insulating state in single crystals of the
  honeycomb lattice material {N}a$_{2}${I}r{O}$_{3}$},}\ }\href {\doibase
  10.1103/PhysRevB.82.064412} {\bibfield  {journal} {\bibinfo  {journal} {Phys.
  Rev. B}\ }\textbf {\bibinfo {volume} {82}},\ \bibinfo {pages} {064412}
  (\bibinfo {year} {2010})}\BibitemShut {NoStop}%
\bibitem [{\citenamefont {Singh}\ \emph {et~al.}(2012)\citenamefont {Singh},
  \citenamefont {Manni}, \citenamefont {Reuther}, \citenamefont {Berlijn},
  \citenamefont {Thomale}, \citenamefont {Ku}, \citenamefont {Trebst},\ and\
  \citenamefont {Gegenwart}}]{PhysRevLett.108.127203}%
  \BibitemOpen
  \bibfield  {author} {\bibinfo {author} {\bibfnamefont {Y.}~\bibnamefont
  {Singh}}, \bibinfo {author} {\bibfnamefont {S.}~\bibnamefont {Manni}},
  \bibinfo {author} {\bibfnamefont {J.}~\bibnamefont {Reuther}}, \bibinfo
  {author} {\bibfnamefont {T.}~\bibnamefont {Berlijn}}, \bibinfo {author}
  {\bibfnamefont {R.}~\bibnamefont {Thomale}}, \bibinfo {author} {\bibfnamefont
  {W.}~\bibnamefont {Ku}}, \bibinfo {author} {\bibfnamefont {S.}~\bibnamefont
  {Trebst}}, \ and\ \bibinfo {author} {\bibfnamefont {P.}~\bibnamefont
  {Gegenwart}},\ }\bibfield  {title} {\enquote {\bibinfo {title} {Relevance of
  the {H}eisenberg-{K}itaev {M}odel for the {H}oneycomb {L}attice {I}ridates
  {${A}_{2}$}{I}r{O}$_{3}$},}\ }\href {\doibase 10.1103/PhysRevLett.108.127203}
  {\bibfield  {journal} {\bibinfo  {journal} {Phys. Rev. Lett.}\ }\textbf
  {\bibinfo {volume} {108}},\ \bibinfo {pages} {127203} (\bibinfo {year}
  {2012})}\BibitemShut {NoStop}%
\bibitem [{\citenamefont {Comin}\ \emph {et~al.}(2012)\citenamefont {Comin},
  \citenamefont {Levy}, \citenamefont {Ludbrook}, \citenamefont {Zhu},
  \citenamefont {Veenstra}, \citenamefont {Rosen}, \citenamefont {Singh},
  \citenamefont {Gegenwart}, \citenamefont {Stricker}, \citenamefont {Hancock},
  \citenamefont {van~der Marel}, \citenamefont {Elfimov},\ and\ \citenamefont
  {Damascelli}}]{PhysRevLett.109.266406}%
  \BibitemOpen
  \bibfield  {author} {\bibinfo {author} {\bibfnamefont {R.}~\bibnamefont
  {Comin}}, \bibinfo {author} {\bibfnamefont {G.}~\bibnamefont {Levy}},
  \bibinfo {author} {\bibfnamefont {B.}~\bibnamefont {Ludbrook}}, \bibinfo
  {author} {\bibfnamefont {Z.-H.}\ \bibnamefont {Zhu}}, \bibinfo {author}
  {\bibfnamefont {C.~N.}\ \bibnamefont {Veenstra}}, \bibinfo {author}
  {\bibfnamefont {J.~A.}\ \bibnamefont {Rosen}}, \bibinfo {author}
  {\bibfnamefont {Y.}~\bibnamefont {Singh}}, \bibinfo {author} {\bibfnamefont
  {P.}~\bibnamefont {Gegenwart}}, \bibinfo {author} {\bibfnamefont
  {D.}~\bibnamefont {Stricker}}, \bibinfo {author} {\bibfnamefont {J.~N.}\
  \bibnamefont {Hancock}}, \bibinfo {author} {\bibfnamefont {D.}~\bibnamefont
  {van~der Marel}}, \bibinfo {author} {\bibfnamefont {I.~S.}\ \bibnamefont
  {Elfimov}}, \ and\ \bibinfo {author} {\bibfnamefont {A.}~\bibnamefont
  {Damascelli}},\ }\bibfield  {title} {\enquote {\bibinfo {title}
  {{N}a$_{2}${I}r{O}$_{3}$ as a {N}ovel {R}elativistic {M}ott {I}nsulator with
  a 340-me{V} gap},}\ }\href {\doibase 10.1103/PhysRevLett.109.266406}
  {\bibfield  {journal} {\bibinfo  {journal} {Phys. Rev. Lett.}\ }\textbf
  {\bibinfo {volume} {109}},\ \bibinfo {pages} {266406} (\bibinfo {year}
  {2012})}\BibitemShut {NoStop}%
\bibitem [{\citenamefont {Modic}\ \emph {et~al.}(2014)\citenamefont {Modic},
  \citenamefont {Smidt}, \citenamefont {Kimchi}, \citenamefont {Breznay},
  \citenamefont {Biffin}, \citenamefont {Choi}, \citenamefont {Johnson},
  \citenamefont {Coldea}, \citenamefont {Watkins-Curry}, \citenamefont
  {McCandless}, \citenamefont {Chan}, \citenamefont {Gandara}, \citenamefont
  {Islam}, \citenamefont {Vishwanath}, \citenamefont {Shekhter}, \citenamefont
  {McDonald},\ and\ \citenamefont {Analytis}}]{modic2014realization}%
  \BibitemOpen
  \bibfield  {author} {\bibinfo {author} {\bibfnamefont {K.~A.}\ \bibnamefont
  {Modic}}, \bibinfo {author} {\bibfnamefont {T.~E.}\ \bibnamefont {Smidt}},
  \bibinfo {author} {\bibfnamefont {I.}~\bibnamefont {Kimchi}}, \bibinfo
  {author} {\bibfnamefont {N.~P.}\ \bibnamefont {Breznay}}, \bibinfo {author}
  {\bibfnamefont {A.}~\bibnamefont {Biffin}}, \bibinfo {author} {\bibfnamefont
  {S.}~\bibnamefont {Choi}}, \bibinfo {author} {\bibfnamefont {R.~D.}\
  \bibnamefont {Johnson}}, \bibinfo {author} {\bibfnamefont {R.}~\bibnamefont
  {Coldea}}, \bibinfo {author} {\bibfnamefont {P.}~\bibnamefont
  {Watkins-Curry}}, \bibinfo {author} {\bibfnamefont {G.~T.}\ \bibnamefont
  {McCandless}}, \bibinfo {author} {\bibfnamefont {J.~Y.}\ \bibnamefont
  {Chan}}, \bibinfo {author} {\bibfnamefont {F.}~\bibnamefont {Gandara}},
  \bibinfo {author} {\bibfnamefont {Z.}~\bibnamefont {Islam}}, \bibinfo
  {author} {\bibfnamefont {A.}~\bibnamefont {Vishwanath}}, \bibinfo {author}
  {\bibfnamefont {A.}~\bibnamefont {Shekhter}}, \bibinfo {author}
  {\bibfnamefont {R.~D.}\ \bibnamefont {McDonald}}, \ and\ \bibinfo {author}
  {\bibfnamefont {J.~G.}\ \bibnamefont {Analytis}},\ }\bibfield  {title}
  {\enquote {\bibinfo {title} {Realization of a three-dimensional
  spin--anisotropic harmonic honeycomb iridate},}\ }\href@noop {} {\bibfield
  {journal} {\bibinfo  {journal} {Nat. Commun.}\ }\textbf {\bibinfo {volume}
  {5}},\ \bibinfo {pages} {4203} (\bibinfo {year} {2014})}\BibitemShut
  {NoStop}%
\bibitem [{\citenamefont {Plumb}\ \emph {et~al.}(2014)\citenamefont {Plumb},
  \citenamefont {Clancy}, \citenamefont {Sandilands}, \citenamefont {Shankar},
  \citenamefont {Hu}, \citenamefont {Burch}, \citenamefont {Kee},\ and\
  \citenamefont {Kim}}]{PhysRevB.90.041112}%
  \BibitemOpen
  \bibfield  {author} {\bibinfo {author} {\bibfnamefont {K.~W.}\ \bibnamefont
  {Plumb}}, \bibinfo {author} {\bibfnamefont {J.~P.}\ \bibnamefont {Clancy}},
  \bibinfo {author} {\bibfnamefont {L.~J.}\ \bibnamefont {Sandilands}},
  \bibinfo {author} {\bibfnamefont {V.~Vijay}\ \bibnamefont {Shankar}},
  \bibinfo {author} {\bibfnamefont {Y.~F.}\ \bibnamefont {Hu}}, \bibinfo
  {author} {\bibfnamefont {K.~S.}\ \bibnamefont {Burch}}, \bibinfo {author}
  {\bibfnamefont {H.-Y.}\ \bibnamefont {Kee}}, \ and\ \bibinfo {author}
  {\bibfnamefont {Y.-J.}\ \bibnamefont {Kim}},\ }\bibfield  {title} {\enquote
  {\bibinfo {title} {$\alpha$-{R}u{C}l$_{3}$: a spin-orbit assisted {M}ott
  insulator on a honeycomb lattice},}\ }\href {\doibase
  10.1103/PhysRevB.90.041112} {\bibfield  {journal} {\bibinfo  {journal} {Phys.
  Rev. B}\ }\textbf {\bibinfo {volume} {90}},\ \bibinfo {pages} {041112}
  (\bibinfo {year} {2014})}\BibitemShut {NoStop}%
\bibitem [{\citenamefont {Kubota}\ \emph {et~al.}(2015)\citenamefont {Kubota},
  \citenamefont {Tanaka}, \citenamefont {Ono}, \citenamefont {Narumi},\ and\
  \citenamefont {Kindo}}]{PhysRevB.91.094422}%
  \BibitemOpen
  \bibfield  {author} {\bibinfo {author} {\bibfnamefont {Y.}~\bibnamefont
  {Kubota}}, \bibinfo {author} {\bibfnamefont {H.}~\bibnamefont {Tanaka}},
  \bibinfo {author} {\bibfnamefont {T.}~\bibnamefont {Ono}}, \bibinfo {author}
  {\bibfnamefont {Y.}~\bibnamefont {Narumi}}, \ and\ \bibinfo {author}
  {\bibfnamefont {K.}~\bibnamefont {Kindo}},\ }\bibfield  {title} {\enquote
  {\bibinfo {title} {Successive magnetic phase transitions in
  $\alpha$-{R}u{C}l$_{3}$: {X}{Y}-like frustrated magnet on the honeycomb
  lattice},}\ }\href {\doibase 10.1103/PhysRevB.91.094422} {\bibfield
  {journal} {\bibinfo  {journal} {Phys. Rev. B}\ }\textbf {\bibinfo {volume}
  {91}},\ \bibinfo {pages} {094422} (\bibinfo {year} {2015})}\BibitemShut
  {NoStop}%
\bibitem [{\citenamefont {Chun}\ \emph {et~al.}(2015)\citenamefont {Chun},
  \citenamefont {Kim}, \citenamefont {Kim}, \citenamefont {Zheng},
  \citenamefont {Stoumpos}, \citenamefont {Malliakas}, \citenamefont
  {Mitchell}, \citenamefont {Mehlawat}, \citenamefont {Singh}, \citenamefont
  {Choi}, \citenamefont {Gog}, \citenamefont {Al-Zein}, \citenamefont {Sala},
  \citenamefont {Krisch}, \citenamefont {Chaloupka}, \citenamefont {Jackeli},
  \citenamefont {Khaliullin},\ and\ \citenamefont {Kim}}]{chun2015direct}%
  \BibitemOpen
  \bibfield  {author} {\bibinfo {author} {\bibfnamefont {S.~H.}\ \bibnamefont
  {Chun}}, \bibinfo {author} {\bibfnamefont {J.-W.}\ \bibnamefont {Kim}},
  \bibinfo {author} {\bibfnamefont {J.}~\bibnamefont {Kim}}, \bibinfo {author}
  {\bibfnamefont {H}~\bibnamefont {Zheng}}, \bibinfo {author} {\bibfnamefont
  {C.~C.}\ \bibnamefont {Stoumpos}}, \bibinfo {author} {\bibfnamefont {C.~D.}\
  \bibnamefont {Malliakas}}, \bibinfo {author} {\bibfnamefont {J.~F.}\
  \bibnamefont {Mitchell}}, \bibinfo {author} {\bibfnamefont {K.}~\bibnamefont
  {Mehlawat}}, \bibinfo {author} {\bibfnamefont {Y.}~\bibnamefont {Singh}},
  \bibinfo {author} {\bibfnamefont {Y}~\bibnamefont {Choi}}, \bibinfo {author}
  {\bibfnamefont {T.}~\bibnamefont {Gog}}, \bibinfo {author} {\bibfnamefont
  {A.}~\bibnamefont {Al-Zein}}, \bibinfo {author} {\bibfnamefont {M.~M.}\
  \bibnamefont {Sala}}, \bibinfo {author} {\bibfnamefont {M.}~\bibnamefont
  {Krisch}}, \bibinfo {author} {\bibfnamefont {J.}~\bibnamefont {Chaloupka}},
  \bibinfo {author} {\bibfnamefont {G.}~\bibnamefont {Jackeli}}, \bibinfo
  {author} {\bibfnamefont {G.}~\bibnamefont {Khaliullin}}, \ and\ \bibinfo
  {author} {\bibfnamefont {B.~J.}\ \bibnamefont {Kim}},\ }\bibfield  {title}
  {\enquote {\bibinfo {title} {Direct evidence for dominant bond-directional
  interactions in a honeycomb lattice iridate {N}a$_2${I}r{O}$_3$},}\
  }\href@noop {} {\bibfield  {journal} {\bibinfo  {journal} {Nat. Phys.}\
  }\textbf {\bibinfo {volume} {11}},\ \bibinfo {pages} {462} (\bibinfo {year}
  {2015})}\BibitemShut {NoStop}%
\bibitem [{\citenamefont {Li}\ \emph {et~al.}(2017)\citenamefont {Li},
  \citenamefont {Li}, \citenamefont {Yu}, \citenamefont {Paramekanti},\ and\
  \citenamefont {Chen}}]{PhysRevB.95.085132}%
  \BibitemOpen
  \bibfield  {author} {\bibinfo {author} {\bibfnamefont {F.-Y.}\ \bibnamefont
  {Li}}, \bibinfo {author} {\bibfnamefont {Y.-D.}\ \bibnamefont {Li}}, \bibinfo
  {author} {\bibfnamefont {Y.}~\bibnamefont {Yu}}, \bibinfo {author}
  {\bibfnamefont {A.}~\bibnamefont {Paramekanti}}, \ and\ \bibinfo {author}
  {\bibfnamefont {G.}~\bibnamefont {Chen}},\ }\bibfield  {title} {\enquote
  {\bibinfo {title} {{K}itaev materials beyond iridates: order by quantum
  disorder and {W}eyl magnons in rare-earth double perovskites},}\ }\href
  {\doibase 10.1103/PhysRevB.95.085132} {\bibfield  {journal} {\bibinfo
  {journal} {Phys. Rev. B}\ }\textbf {\bibinfo {volume} {95}},\ \bibinfo
  {pages} {085132} (\bibinfo {year} {2017})}\BibitemShut {NoStop}%
\bibitem [{\citenamefont {Griffith}(1961)}]{griffith1961theory}%
  \BibitemOpen
  \bibfield  {author} {\bibinfo {author} {\bibfnamefont {J.~S.}\ \bibnamefont
  {Griffith}},\ }\href@noop {} {\emph {\bibinfo {title} {The theory of
  transition-metal ions}}}\ (\bibinfo  {publisher} {Cambridge University
  Press},\ \bibinfo {year} {1961})\BibitemShut {NoStop}%
\bibitem [{\citenamefont {Sugano}(1970)}]{sugano2012multiplets}%
  \BibitemOpen
  \bibfield  {author} {\bibinfo {author} {\bibfnamefont {S.}~\bibnamefont
  {Sugano}},\ }\href@noop {} {\emph {\bibinfo {title} {Multiplets of
  transition-metal ions in crystals}}}\ (\bibinfo  {publisher} {Elsevier},\
  \bibinfo {year} {1970})\BibitemShut {NoStop}%
\bibitem [{\citenamefont {Slater}\ and\ \citenamefont
  {Koster}(1954)}]{PhysRev.94.1498}%
  \BibitemOpen
  \bibfield  {author} {\bibinfo {author} {\bibfnamefont {J.~C.}\ \bibnamefont
  {Slater}}\ and\ \bibinfo {author} {\bibfnamefont {G.~F.}\ \bibnamefont
  {Koster}},\ }\bibfield  {title} {\enquote {\bibinfo {title} {Simplified
  {L}{C}{A}{O} {M}ethod for the {P}eriodic {P}otential {P}roblem},}\ }\href
  {\doibase 10.1103/PhysRev.94.1498} {\bibfield  {journal} {\bibinfo  {journal}
  {Phys. Rev.}\ }\textbf {\bibinfo {volume} {94}},\ \bibinfo {pages} {1498}
  (\bibinfo {year} {1954})}\BibitemShut {NoStop}%
\bibitem [{\citenamefont {Harrison}(1980)}]{harrison2012electronic}%
  \BibitemOpen
  \bibfield  {author} {\bibinfo {author} {\bibfnamefont {W.~A.}\ \bibnamefont
  {Harrison}},\ }\href@noop {} {\emph {\bibinfo {title} {Electronic structure
  and the properties of solids: the physics of the chemical bond}}}\ (\bibinfo
  {publisher} {Courier Corporation},\ \bibinfo {year} {1980})\BibitemShut
  {NoStop}%
\bibitem [{\citenamefont {Nasu}\ \emph {et~al.}(2015)\citenamefont {Nasu},
  \citenamefont {Udagawa},\ and\ \citenamefont {Motome}}]{PhysRevB.92.115122}%
  \BibitemOpen
  \bibfield  {author} {\bibinfo {author} {\bibfnamefont {J.}~\bibnamefont
  {Nasu}}, \bibinfo {author} {\bibfnamefont {M.}~\bibnamefont {Udagawa}}, \
  and\ \bibinfo {author} {\bibfnamefont {Y.}~\bibnamefont {Motome}},\
  }\bibfield  {title} {\enquote {\bibinfo {title} {Thermal fractionalization of
  quantum spins in a {K}itaev model: {T}emperature-linear specific heat and
  coherent transport of {M}ajorana fermions},}\ }\href {\doibase
  10.1103/PhysRevB.92.115122} {\bibfield  {journal} {\bibinfo  {journal} {Phys.
  Rev. B}\ }\textbf {\bibinfo {volume} {92}},\ \bibinfo {pages} {115122}
  (\bibinfo {year} {2015})}\BibitemShut {NoStop}%
\bibitem [{\citenamefont {Liu}\ and\ \citenamefont
  {Khaliullin}(2017)}]{liu2017}%
  \BibitemOpen
  \bibfield  {author} {\bibinfo {author} {\bibfnamefont {H.}~\bibnamefont
  {Liu}}\ and\ \bibinfo {author} {\bibfnamefont {G.}~\bibnamefont
  {Khaliullin}},\ }\href@noop {} {\enquote {\bibinfo {title} {Pseudospin
  exchange interactions in $d^7$ cobalt compounds: possible realization of the
  {K}itaev model},}\ } (\bibinfo {year} {2017}),\ \Eprint
  {http://arxiv.org/abs/arXiv:1710.10193} {arXiv:1710.10193} \BibitemShut
  {NoStop}%
\end{thebibliography}%
\end{document}